\documentclass[preprint,prd,aps,showpacs,showkeys,onecolumn,english]{revtex4}
\usepackage{amsmath,amssymb} 
\usepackage[dvips]{graphicx} 
\usepackage{graphics}
\usepackage{graphicx}
\usepackage{epsfig}
\usepackage{slashed}
\usepackage[english]{babel}

\newcommand\nn{\nonumber}
\newcommand\ba{\begin{eqnarray}}
\newcommand\ea{\end{eqnarray}}

\begin{document}
\title{The calculation of differential and total cross sections for $W^+ W^- \gamma$ production process in proton-proton collisions at LHC energies}
\author{Azad~I.~Ahmadov$^{a,b}$ \footnote{E-mail: ahmadov@theor.jinr.ru}}
\affiliation{$^{a}$ Bogoliubov Laboratory of Theoretical Physics,
JINR, Dubna, 141980 Russia}
\affiliation{$^{b}$ Institute of Physics, Azerbaijan
National Academy of Sciences, H.Javid ave. 131, AZ-1143 Baku, Azerbaijan}

\date{\today}

\begin{abstract}
The vector bosons production at the Large Hadron Collider makes it possible to investigate in detail the
basic structure of electroweak interactions.
Besides the LHC with a good accuracy will be measure production of weak bosons ($pp \to  W^+W^-$).
The production of weak bosons with photon ($pp \to  W^+W^-\gamma$) provides an increasingly
powerful handle at higher center-of-mass energies. \\
We present phenomenological results for $WW\gamma$ production in proton-proton interaction at the Large Hadron Collider.
In this paper, we calculate the total and differential cross sections.
We consider the dependence of differential cross section distributions on transverse momentum and rapidity particles,
which are produced in the final state ($W^+$ and $W^-$).
We consider several important distributions, which are included in the search for new physics at the Large Hadron Collider. \\
The results for transverse momentum distributions, rapidity distributions and total cross section are presented.

\vspace*{0.5cm}

\noindent
\end{abstract}

\maketitle

\section{Introduction}
\label{Introduction}
The fundamental description of matter and the forces that determine its behavior
are constantly studied in particle physics. \\
The new physics, which underlies the dynamics of the electroweak symmetry breaking is one of the outstanding
open questions of the Standard Model. \\
Massive vector boson production is among the most important electroweak processes at hadron colliders.
This makes is it possible to study in detail the structure of the gauge symmetry of electroweak interactions and the electroweak symmetry breaking mechanism. \\
At Tevatron and the LHC, various measurements of the $W^+ W^-$ production have been carried out \cite{I1,I2,II1,II2,II3,II4,II5,I3,IC1,IC2,IC3,IC4,I4,I5,I6}
and it was shown that the total $W^+W^-$ cross section at 8 $TeV$ and 13 $TeV$ exceeds theoretical expectations and causes possible phenomena of new physics \cite{I71,I72,I7,I8,I9,I10,I11,I12,I13}. \\
The vector-bosons production plays an important role in various areas of the Large Hadron Collider (LHC) physics programme.
Experimental studies of these processes permit one to test key aspects of the Standard Model (SM) at energies up to the TeV regime.

It should be noted that precise calculations of vector-bosons production have reached a new era with the recent results at the
next-to-next-to-leading order (NNLO) QCD precision for V + 1 jet \cite{L1} as well as have been used to precisely describe backgrounds
for dark matter searches \cite{L2}.  \\
In recent years, owing to the Large Hadron Collider (LHC), particle physicists at CERN have gained access to more accurate
testing of the parameters of the Standard Model.  \\
Therefore, we want to consider an important SM process of the vector boson production in proton-proton collisions at LHC energy. \\
At leading order of perturbation theory, we will consider the $W^+W^-\gamma$ production described by one of the subprocesses,
namely, quark-antiquark scattering. \\
At present, one of the prime targets for experiments is the measurement of the $W W \gamma$ and $W W Z$ couplings.
In the Standard Model these couplings are unambiguously fixed by the non-abelian nature of the SU (2) $\times$ U (1) gauge symmetry.

The underlying theory of our calculations is the SM of particle physics.
Since we deal with the production of weak bosons, we will take into account the electroweak sector, too. \\
One of the functions that describes the interaction of partons in a hadron is the distribution of
the dependence on the transverse momentum.
The transverse momentum dependent distributions naturally appear within factorization theorem for the
differential cross section of inclusive hard processes \cite{Col1,Col2}. \\
The Large Hadron Collider (LHC) at CERN allows the vector boson production with transverse momenta in the TeV regime
in proton-proton collisions (pp) with a centre-of-mass energy of $\sqrt {s} = 7 \,\,{\mbox {or}}\,\, 8 \,\,{\mbox {or}}\,\, 14\,\, TeV$. \\
The production $WW\gamma$ and $WZ\gamma$ of triboson has been studied from proton-proton collisions at a centre-of-mass energy of $\sqrt {s} =8 \,\,TeV$
recorded with the ATLAS detector corresponding to an integrated luminosity of 20.2 $fb^{-1}$ \cite{ATLAS}, and the CMS detector data correspond to an integrated luminosity
of 19.3 $fb^{-1}$ \cite{CMS} at the LHC. \\
The $WW\gamma$ production in the proton-proton collision is analyzed in the \cite{Sokolenko}, and
it is shown that this channel can discover new physics at the LHC. \\
In some works \cite{Y,J,B} using ATLAS and CMS  data, the di- and multiboson $(ZZ, WZ, WW, WW\gamma \,\,\,{\rm and}\,\,\, WZ\gamma)$ productions $pp$ collisions
at $\sqrt {s} = 8 \,\,\,{\rm and}\,\,\, 13 \,\,TeV$ are studied. \\
The detailed theoretical investigation was carried out of the $WW, WZ, ZZ, W\gamma, Z\gamma$ vector bosons production and $WWg, ZZg, ZZq$ productions
in the process of $pp$ collisions \cite{WW1,WW2,WW3,WW4,WW41,WW42}.
In \cite{WW5}, was theoretically  investigated the production of $W^+W^-\gamma$ in the process of $pp$ collisions. \\
It should be noted that the study of the $W^+W^-\gamma$ production process may be will be for the good perform of accurate tests of the description
of electroweak and strong interactions in the standard model (SM). \\
The Standard Model of particle physics describes most of high-energy experimental data.
In \cite{SM1}, the cross-section of the $WW$ and $WZ$ production has been precisely measured at the LHC and found to be in agreement with the SM expectation.
By extending the analysis of the inclusive $WW+WZ$ diboson production cross section in proton-proton collisions,
the production of three gauge bosons $WW (WZ) \gamma$ \cite{SM2} was observed.  \\
It can be predicted that the $WW\gamma$ production also provides a potential background for new physics searches, such as supersymmetric particles.  \\
There is great interest in the two and three gauge-boson production of final states in proton-proton interaction at high-energy colliders,
since they will allow crucial tests of the electroweak gauge theory. \\
Taking into account the above, one can come to the conclusion that the study of the $WW\gamma$ production in proton-proton collisions presents
the most interesting problems for LHC and other colliders. \\
Thus, in the present paper we will investigate these problems; i.e. considering the LHC conditions, we will calculate cross sections of the $WW\gamma$
production in proton-proton collisions.

We can write convolution of the matrix element of process and the universal parton density functions for
obtaining an inclusive cross section for a given scattering process with the use of the factorization theorem in perturbative quantum chromodynamics (pQCD). \\
We present the cross sections for the vector boson production to study the effect of contributions separating single boson and diboson production.

\section{General framework \label{ht}}

In this section we want to discuss the $WW\gamma$ three bosons production in hadron-hadron collisions at LHC energy. \\
The process is written in the form

\ba
p(P_1) + p(P_2) \to W^-(k_1) + W^+(k_2) + \gamma (k_3).
\label{A1}
\ea
The $W^+W^-\gamma$ production at the LHC is obtained form the quark-antiquark annihilation subprocess at the parton level:
\ba
q(p_1) + \bar {q}(p_2) \to W^-(k_1) + W^+(k_2) + \gamma (k_3).
\label{A2}
\ea
To calculate the cross section, we need to consider all diagrams. In this process, we have fifteen Feynman diagrams.  \\
The leading order (LO) of all Feynman diagrams for \eqref{A2} process is illustrated in Figs.1, 2 and 3, respectively.
\begin{figure}[!htb]
       \centering
       \includegraphics[width=0.71\linewidth]{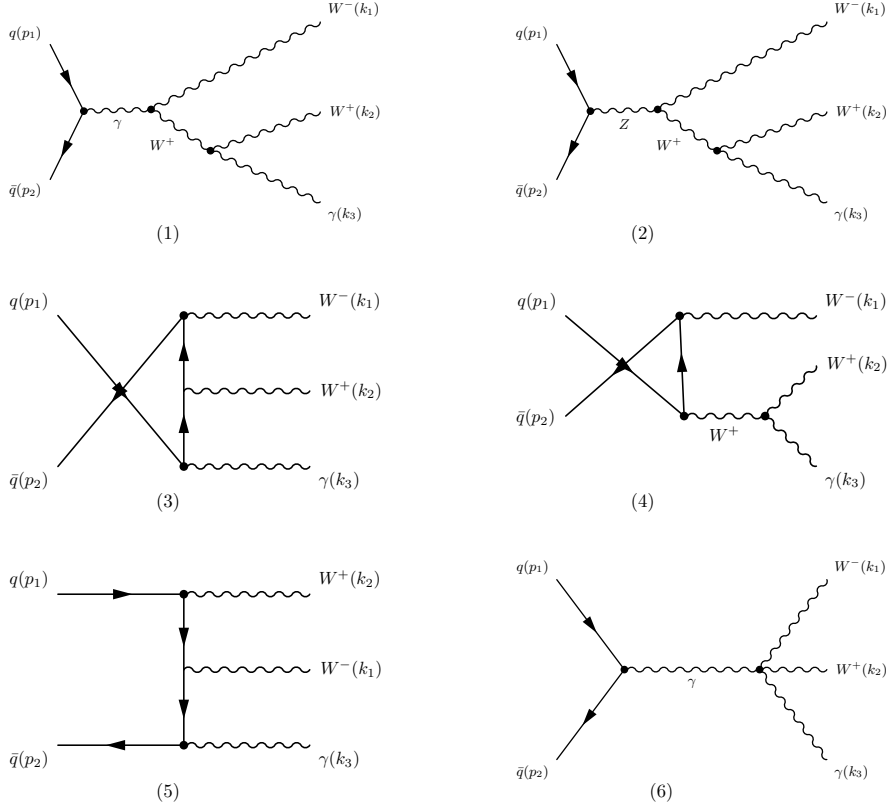}
       \caption{The Feynman diagrams for the process $q \bar {q} \to W^+W^-\gamma$.}
       \label{diagram1}
\end{figure}
\begin{figure}[!htb]
       \centering
       \includegraphics[width=0.71\linewidth]{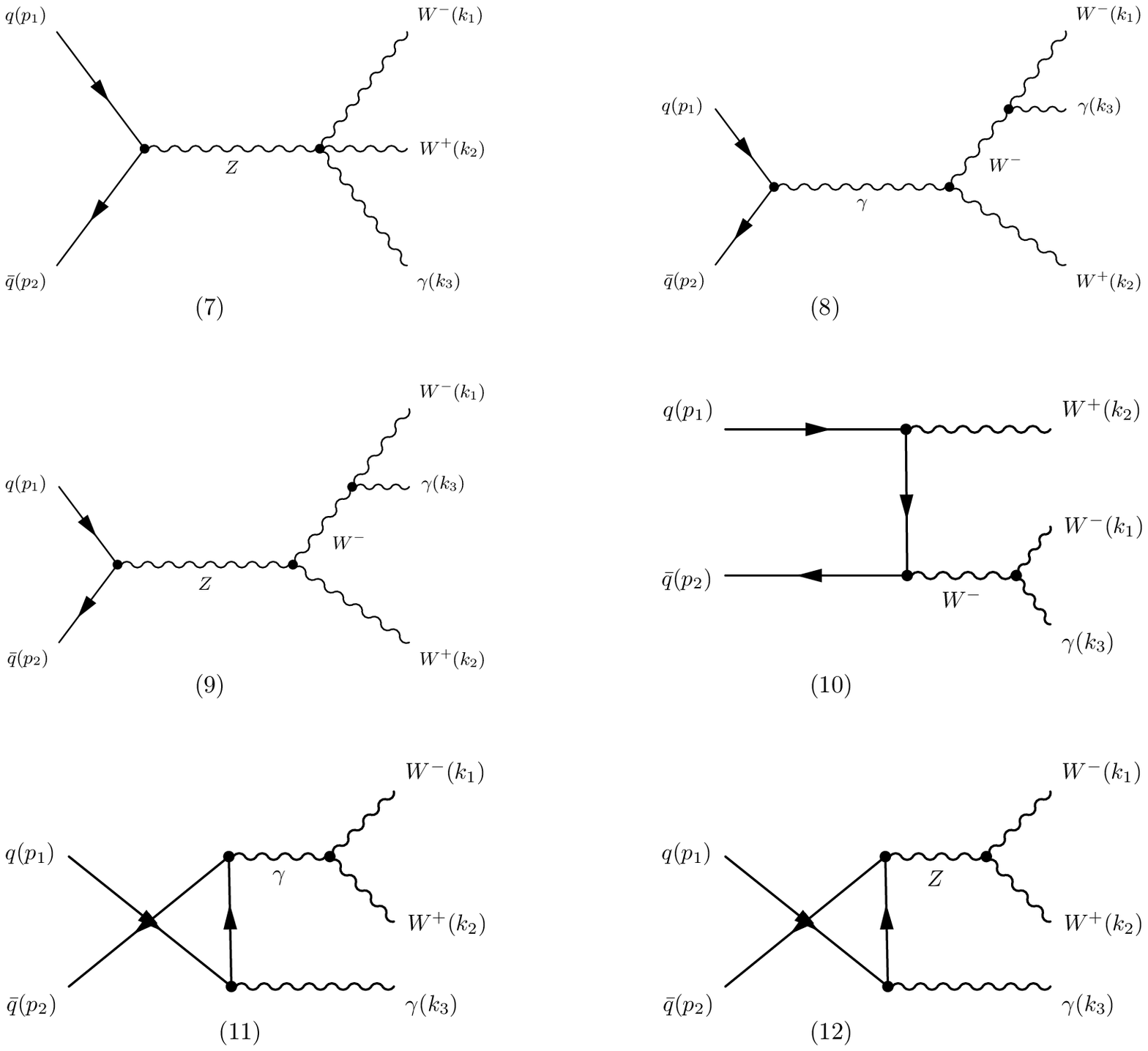}
       \caption{The Feynman diagrams for the process $q \bar {q} \to W^+W^-\gamma$.}
        \label{diagram2}
\end{figure}
\begin{figure}[!htb]
       \centering
       \includegraphics[width=0.71\linewidth]{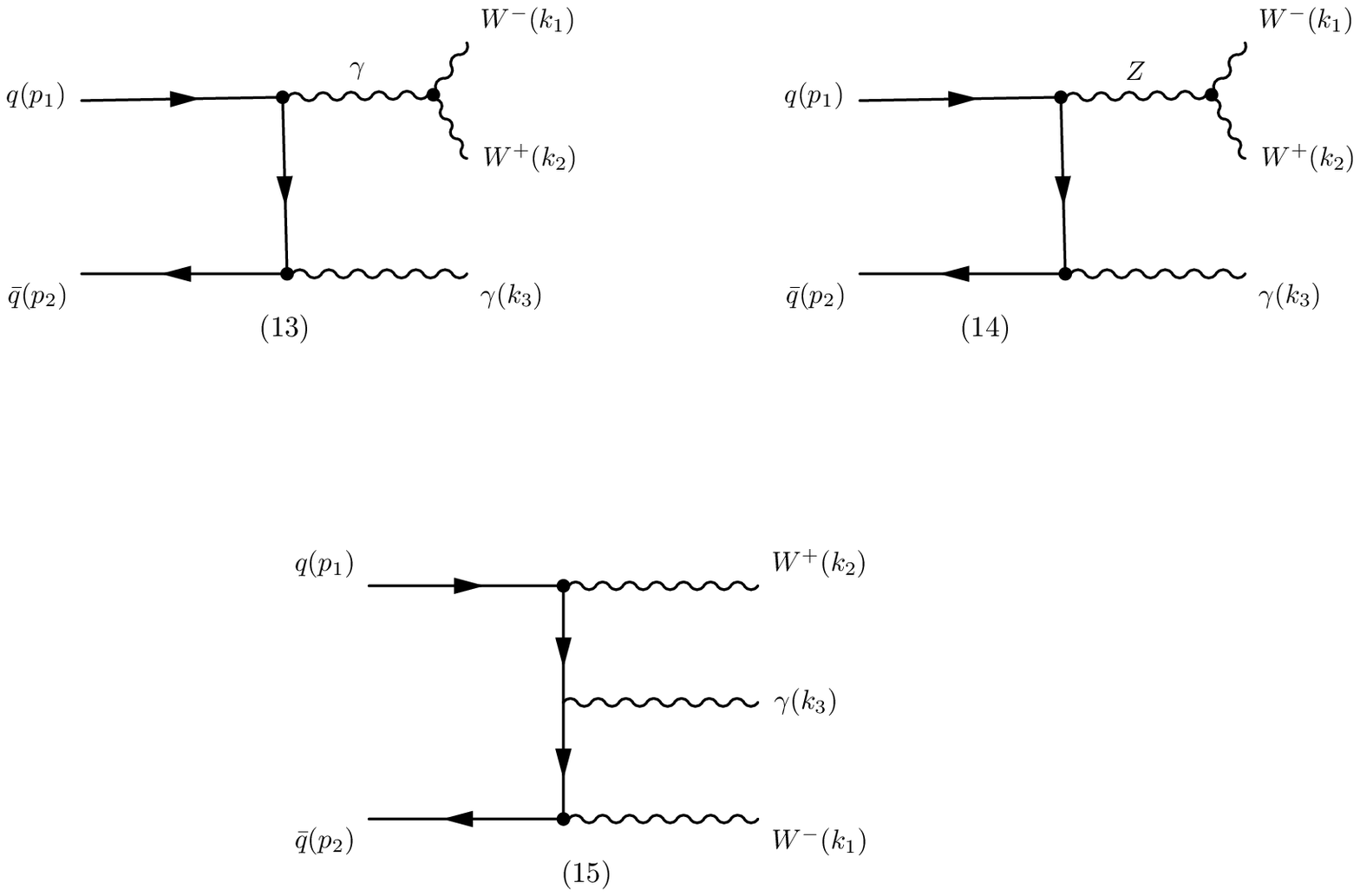}
       \caption{The Feynman diagrams for the process $q \bar {q} \to W^+W^-\gamma$.}
        \label{diagram3}
\end{figure}
%

The general couplings of two charged vector bosons with a neutral vector boson,
$WW\gamma$ and $WWZ$, can be derived from the following effective Lagrangian, which conserves $C$ (charge) and $P$ (parity) separately, can be given
as follows \cite{L3, L4, L5}:
\begin{eqnarray}
{\mathcal L} & = & ig_{WW\gamma}[g_{1}^{\gamma}(W_{\mu\nu}^{\dagger}W^{\mu}A^{\nu}-W^{\mu\nu}W_{\mu}^{\dagger}A_{\nu})+
\kappa_{\gamma}W_{\mu}^{\dagger}W_{\nu}A^{\mu\nu}+\frac{\lambda_{\gamma}}{m_{W}^{2}}W_{\rho\mu}^{\dagger}W_{\nu}^{\mu}A^{\nu\rho}]\nonumber \\
& + & ig_{WWZ}[g_{1}^{Z}(W_{\mu\nu}^{\dagger}W^{\mu}Z^{\nu}-W^{\mu\nu}W_{\mu}^{\dagger}Z_{\nu})+\kappa_{Z}W_{\mu}^{\dagger}W_{\nu}Z^{\mu\nu}+
\frac{\lambda_{Z}}{m_{W}^{2}}W_{\rho\mu}^{\dagger}W_{\nu}^{\mu}Z^{\nu\rho}],
\label{AL}
\end{eqnarray}
where $X_{\mu\nu} \equiv \partial_{\mu}X_{\nu} - \partial_{\nu}X_{\mu}$ $(X_{\mu} =W_{\mu}, A_{\mu}, Z_{\mu}$) and the coefficients $g^{\gamma, Z}_1$, $k_{\gamma, Z}$ and $\lambda_{\gamma, Z}$ are defined to be
$g^{\gamma, Z}_1 = k_{\gamma, Z} = 1$ and $\lambda_{\gamma, Z} = 0$ in the Standard Model, and the overall coupling constants $g_{WWV}$ are given by
$g_{WW\gamma} =-e$ and $g_{WWZ} = -e \cot\theta_W$, respectively, with $\theta_W$ being the weak mixing angle and $e$ being the positron charge. \\
It should be noted that triple gauge couplings were thoroughly measured at LEP2 \cite{LEP}. \\
For the W W interaction, many phenomenological processes at linear and hadron colliders were studied \cite{LW1,LW2,LW3,LW4,LW5,LW6,LW7,LW8,LW9,LW10}.

We can write the Feynman amplitudes for the partonic process $q + \bar {q} \to W^- + W^+ + \gamma$ as

\clearpage

\ba
M_1 &=& -K_1 \cdot \bar{v}(p_2,m_q)\cdot \gamma_{\mu} \cdot u(p_1,m_q)\cdot \biggl[(k_2 - k_3)^{\rho}g_{\beta\nu} + (-2k_2 - k_3)^{\nu}g_{\beta\rho}+
(k_2 + 2k_3)^{\beta}g_{\nu\rho}\biggr] \cdot \nn \\
&&\cdot g_{\mu\sigma}\cdot \biggl[(2k_1 +k_2+k_3)^{\lambda}g_{\alpha\sigma} +(k_2+k_3-k_1)^{\sigma}g_{\alpha\lambda} +
(-k_1-2k_2-2k_3)^{\alpha}g_{\sigma\lambda}\biggr] \cdot \nn \\
&&\cdot g_{\rho\lambda}\cdot \frac{1}{(k_2+k_3)^2-M^2_W +i M_W \Gamma_W}\cdot \frac{1}{(k_1+k_2+k_3)^2} \cdot
\varepsilon^{\ast}_{\nu}(k_3) \varepsilon^{\ast}_{\alpha}(k_1) \varepsilon^{\ast}_{\beta}(k_2); \nn \\
M_2 &=& K_2 \cdot \bar{v}(p_2,m_q) \cdot \gamma_{\mu} \cdot u(p_1,m_q) \cdot \biggl[(k_2-k_3)^{\rho}g_{\beta\nu}+(-2k_2-k_3)^{\nu}g_{\beta\rho}+
(k_2+2k_3)^{\beta}g_{\nu\rho}\biggr] \cdot \nn \\
&&\cdot g_{\mu\sigma}\cdot \biggl[(2k_1 +k_2+k_3)^{\lambda}g_{\alpha\sigma} +(k_2+k_3-k_1)^{\sigma}g_{\alpha\lambda} +
(-k_1-2k_2-2k_3)^{\alpha}g_{\sigma\lambda}\biggr] \cdot \nn \\
&&\cdot g_{\rho\lambda}\cdot \frac{1}{(k_2+k_3)^2-M^2_W+i M_W \Gamma_W}\cdot \frac{1}{(k_1+k_2+k_3)^2-M^2_Z+i M_Z \Gamma_Z} \cdot
\varepsilon^{\ast}_{\nu}(k_3) \varepsilon^{\ast}_{\alpha}(k_1) \varepsilon^{\ast}_{\beta}(k_2); \nn \\
M_3 &=& -K_3 \cdot \bar{v}(p_2,m_q) \cdot \gamma_{\mu} (\hat{k}_1-\hat{p}_2 +m_q) \gamma_{\nu}
(\hat{k}_1+\hat{k}_2-\hat{p}_2 +m_q)\gamma_{\rho} \cdot u(p_1,m_q)\cdot \nn \\
&&\cdot \frac{1}{(p_2-k_1)^2-m^2_q}\cdot \frac{1}{(p_2-k_1-k_2)^2-m^2_q} \cdot \varepsilon^{\ast}_{\rho}(k_3) \varepsilon^{\ast}_{\mu}(k_1) \varepsilon^{\ast}_{\nu}(k_2); \nn \\
M_4 &=& -K_4 \cdot \bar{v}(p_2,m_q) \cdot \gamma_{\mu} (\hat{k}_1-\hat{p}_2 +m_q)\gamma_{\nu} \cdot u(p_1,m_q) \cdot \biggl[(k_2-k_3)^{\rho}g_{\sigma\alpha}+(-2k_2-k_3)^{\alpha}g_{\sigma\rho}+ \nn \\
&&+ (k_2+2k_3)^{\sigma}g_{\rho\alpha}\biggr]g_{\nu\rho}\cdot \frac{1}{(p_2-k_1)^2-m^2_q}\cdot \frac{1}{(k_2+k_3)^2-M^2_W +i M_W \Gamma_W} \cdot
\varepsilon^{\ast}_{\alpha}(k_3) \varepsilon^{\ast}_{\mu}(k_1) \varepsilon^{\ast}_{\sigma}(k_2); \nn \\
M_5 &=& -K_5 \cdot \bar{v}(p_2,m_q) \cdot \gamma_{\mu} (\hat{k}_3-\hat{p}_2 +m_q) \gamma_{\nu}
(\hat{k}_1+\hat{k}_3-\hat{p}_2 +m_q)\gamma_{\rho} \cdot u(p_1,m_q)\cdot \nn \\
&&\cdot \frac{1}{(p_2-k_3)^2-m^2_q}\cdot \frac{1}{(p_2-k_1-k_3)^2-m^2_q} \cdot
\varepsilon^{\ast}_{\mu}(k_3) \varepsilon^{\ast}_{\nu}(k_1) \varepsilon^{\ast}_{\rho}(k_2); \nn \\
M_6 &=& K_6\cdot \bar{v}(p_2,m_q) \cdot \gamma_{\mu} \cdot u(p_1,m_q)\cdot g_{\mu\sigma}
\cdot \frac{1}{(k_1+k_2+k_3)^2} \cdot
\varepsilon^{\ast}_{\alpha}(k_1) \varepsilon^{\ast}_{\beta}(k_2) \varepsilon^{\ast}_{\nu}(k_3)\cdot \nn \\
&&\cdot [g_{\alpha\sigma}g_{\beta\nu}+g_{\alpha\nu}g_{\beta\sigma}-2g_{\alpha\beta}g_{\nu\sigma}]; \nn \\
M_7 &=& -K_7 \cdot \bar{v}(p_2,m_q) \cdot \gamma_{\mu} \cdot u(p_1,m_q) \cdot g_{\mu\sigma}\frac{1}{(k_1+k_2+k_3)^2-M^2_Z+i M_Z \Gamma_Z} \cdot \nn \\
&&\cdot \varepsilon^{\ast}_{\alpha}(k_1) \varepsilon^{\ast}_{\beta}(k_2) \varepsilon^{\ast}_{\nu}(k_3)
\cdot \biggl[-g_{\alpha\sigma}g_{\beta\nu} -g_{\alpha\nu}g_{\beta\sigma} + 2g_{\alpha\beta}g_{\nu\sigma}\biggr];  \nn \\
M_8 &=& K_8 \cdot \bar{v}(p_2,m_q) \cdot \gamma_{\mu} \cdot u(p_1,m_q) \cdot \biggl[(k_1-k_3)^{\rho}g_{\alpha\nu}+(-2k_1-k_3)^{\nu}g_{\alpha\rho}+
(k_1+2k_3)^{\alpha}g_{\nu\rho}\biggr]g_{\mu\sigma}\cdot \nn \\
&&\cdot \biggl[(k_1 +2k_2+k_3)^{\lambda}g_{\beta\sigma} +(k_1-k_2+k_3)^{\sigma}g_{\beta\lambda} +
(-2k_1-k_2-2k_3)^{\beta}g_{\sigma\lambda}\biggr]g_{\rho\lambda}\cdot  \nn \\
&&\cdot \frac{1}{(k_1+k_3)^2-M^2_W+i M_W \Gamma_W}\cdot \frac{1}{(k_1+k_2+k_3)^2} \cdot
\varepsilon^{\ast}_{\nu}(k_3) \varepsilon^{\ast}_{\alpha}(k_1) \varepsilon^{\ast}_{\beta}(k_2); \nn
\ea
\ba
M_9 &=& K_9 \cdot \bar{v}(p_2,m_q) \cdot \gamma_{\mu}  \cdot u(p_1,m_q) \cdot \biggl[(k_1-k_3)^{\rho}g_{\alpha\nu}+(-2k_1-k_3)^{\nu}g_{\alpha\rho}+
(k_1+2k_3)^{\alpha}g_{\nu\rho}\biggr]g_{\mu\sigma}\cdot \nn \\
&&\cdot \biggl[(k_1 +2k_2+k_3)^{\lambda}g_{\beta\sigma} +(k_1-k_2+k_3)^{\sigma}g_{\beta\lambda} +
(-2k_1-k_2-2k_3)^{\beta}g_{\sigma\lambda}\biggr]g_{\rho\lambda}\cdot \nn \\
&&\cdot \frac{1}{(k_1+k_3)^2-M^2_W+i M_W \Gamma_W}\cdot \frac{1}{(k_1+k_2+k_3)^2-M^2_Z+i M_Z \Gamma_Z} \cdot
\varepsilon^{\ast}_{\nu}(k_3) \varepsilon^{\ast}_{\alpha}(k_1) \varepsilon^{\ast}_{\beta}(k_2); \nn \\
M_{10} &=& -K_{10} \cdot \bar{v}(p_2,m_q) \cdot \gamma_{\mu} (\hat{k}_1 +\hat{k}_3 -\hat{p}_2 +m_q)\gamma_{\nu} \cdot u(p_1,m_q) \cdot \biggl[(k_1-k_3)^{\alpha}g_{\sigma\rho}+(-2k_1-k_3)^{\sigma}g_{\alpha\rho}+ \nn \\
&&+(k_1+2k_3)^{\rho}g_{\sigma\alpha}\biggr]g_{\nu\alpha}\cdot \frac{1}{(p_2-k_1-k_3)^2-m^2_q}\cdot \frac{1}{(k_1+k_3)^2-M^2_W+i M_W \Gamma_W} \cdot \nn \\
&&\cdot \varepsilon^{\ast}_{\sigma}(k_3) \varepsilon^{\ast}_{\rho}(k_1) \varepsilon^{\ast}_{\nu}(k_2);  \nn \\
M_{11} &=& -K_{11} \cdot \bar{v}(p_2,m_q) \cdot \gamma_{\mu} (\hat{k}_1 +\hat{k}_2 -\hat{p}_2 +m_q)\gamma_{\nu} \cdot u(p_1,m_q) \cdot \biggl[(k_1-k_2)^{\sigma}g_{\alpha\beta}+(-2k_1-k_2)^{\beta}g_{\alpha\sigma}+ \nn \\
&&+ (k_1+2k_2)^{\alpha}g_{\beta\sigma}\biggr]g_{\mu\sigma}\cdot \frac{1}{(p_2-k_1-k_2)^2-m^2_q}\cdot \frac{1}{(k_1+k_2)^2} \cdot
\varepsilon^{\ast}_{\nu}(k_3) \varepsilon^{\ast}_{\alpha}(k_1) \varepsilon^{\ast}_{\beta}(k_2); \nn \\
M_{12} &=& K_{12} \cdot \bar{v}(p_2,m_q) \cdot \gamma_{\mu} (\hat{k}_1 +\hat{k}_2 -\hat{p}_2 +m_q) \gamma_{\nu} \cdot u(p_1,m_q)\cdot \biggl[(k_1-k_2)^{\sigma}g_{\alpha\beta}+(-2k_1-k_2)^{\beta}g_{\alpha\sigma}+ \nn \\
&&+ (k_1+2k_2)^{\alpha}g_{\beta\sigma}\biggr]g_{\mu\sigma}\cdot \frac{1}{(p_2-k_1-k_2)^2-m^2_q}\cdot \frac{1}{(k_1+k_2)^2-M^2_Z+i M_Z \Gamma_Z} \cdot
\varepsilon^{\ast}_{\nu}(k_3) \varepsilon^{\ast}_{\alpha}(k_1) \varepsilon^{\ast}_{\beta}(k_2);  \nn \\
M_{13} &=& K_{13} \cdot \bar{v}(p_2,m_q) \cdot \gamma_{\mu} (\hat{k}_3 -\hat{p}_2 +m_q)\gamma_{\nu} \cdot u(p_1,m_q) \cdot \biggl[(k_1-k_2)^{\sigma}g_{\alpha\beta}+(-2k_1-k_2)^{\beta}g_{\alpha\sigma}+ \nn \\
&&+ (k_1+2k_2)^{\alpha}g_{\beta\sigma}\biggr]g_{\nu\sigma}\cdot \frac{1}{(p_2-k_3)^2-m^2_q}\cdot \frac{1}{(k_1+k_2)^2} \cdot
\varepsilon^{\ast}_{\mu}(k_3) \varepsilon^{\ast}_{\alpha}(k_1) \varepsilon^{\ast}_{\beta}(k_2); \nn \\
M_{14} &=& K_{14} \cdot \bar{v}(p_2,m_q) \cdot \gamma_{\mu} (\hat{k}_3 -\hat{p}_2 +m_q) \gamma_{\nu}  \cdot u(p_1,m_q)\cdot \biggl[(k_1-k_2)^{\sigma}g_{\alpha\beta}+(-2k_1-k_2)^{\beta}g_{\alpha\sigma}+  \nn \\
&&+ (k_1+2k_2)^{\alpha}g_{\beta\sigma}\biggr]g_{\nu\sigma}\cdot \frac{1}{(p_2-k_3)^2-m^2_q}\cdot \frac{1}{(k_1+k_2)^2-M^2_Z+i M_Z \Gamma_Z} \cdot
\varepsilon^{\ast}_{\mu}(k_3) \varepsilon^{\ast}_{\alpha}(k_1) \varepsilon^{\ast}_{\beta}(k_2); \nn \\
M_{15} &=& K_{15} \cdot \bar{v}(p_2,m_q) \cdot \gamma_{\mu} (\hat{k}_1-\hat{p}_2 +m_q) \gamma_{\nu}
(\hat{k}_1+\hat{k}_3-\hat{p}_2 +m_q)\gamma_{\rho} \cdot u(p_1,m_q)\cdot \nn \\
&&\cdot \frac{1}{(p_2-k_1)^2-m^2_q}\cdot \frac{1}{(p_2-k_1-k_3)^2-m^2_q} \cdot
\varepsilon^{\ast}_{\nu}(k_3) \varepsilon^{\ast}_{\mu}(k_1) \varepsilon^{\ast}_{\rho}(k_2);
\label{A3}
\ea
here $M_W$, $M_Z$, $\Gamma_W$ and $\Gamma_Z$ are the masses and the decay width of the $W$ and $Z$ bosons, respectively.
$p_1$ and $p_2$ are the 4-momenta of the quark and anti-quark in the initial state.
$\varepsilon_{\mu}(k_1),\,\,\,\varepsilon_{\nu}(k_2)$ are the polarization vectors of $W^-, \,\,W^+$ bosons, and $\varepsilon_{\rho}(k_3)$ is the polarization vector of the photon,
$k_1,\,\,\,k_2$, and $k_3$ are 4-momenta of $W^-,\,\,W^+$, and photon, respectively. \\
$K_1, K_2, K_3, K_4, K_5, K_6, K_7, K_8, K_9, K_{10}, K_{11}, K_{12}, K_{13}, K_{14}, K_{15}$ are the coefficients that are obtained from the Feynman rule for every amplitude of the diagrams, respectively.
\ba
K_1 = \frac{4e^3}{3};\,\,\,\,
K_2 = e^3 \cdot \frac{Cos\theta_W}{Sin\theta_W} \cdot
 \biggl[\frac{\frac{1}{2}-\frac{2Sin^2\theta_{W}}{3}}{Cos\theta_W Sin\theta_W} -\frac{2Sin\theta_W}{3Cos\theta_W} \biggr]; \,\,\,\,
 K_3 = \frac{16e}{3}M^2_W \frac{G_F}{\sqrt {2}} |V_{ud}|^2; \nn \\
 K_4 = 4e M^2_W \frac{G_F}{\sqrt {2}} |V_{ud}|^2; \,\,\,\,
 K_5 = \frac{16e}{3}  M^2_W \frac{G_F}{\sqrt {2}} |V_{ud}|^2; \,\,\,\,
 K_6 = \frac{4e^3}{3}; \nn \\
 K_7 =  e^3\frac{Cos\theta_W}{Sin\theta_W} \cdot \biggl[\frac{\frac{1}{2}-\frac{2Sin^2\theta_{W}}{3}}{Cos\theta_W Sin\theta_W} -
\frac{2Sin\theta_W}{3Cos\theta_W} \biggr];  \,\,\,\,
K_8 = \frac{4e^3}{3}; \nn \\
 K_9 =  e^3\frac{Cos\theta_W}{Sin\theta_W} \cdot \biggl[\frac{\frac{1}{2}-\frac{2Sin^2\theta_{W}}{3}}{Cos\theta_W Sin\theta_W} -
\frac{2Sin\theta_W}{3Cos\theta_W} \biggr]; \,\,\,\,
K_{10} = 4e  M^2_W \frac{G_F}{\sqrt {2}} |V_{ud}|^2; \,\,\,\,
K_{11} = \frac{16e^3}{9}; \nn \\
K_{12} =   \frac{4e^3}{3}\cdot \frac{Cos\theta_W}{Sin\theta_W} \cdot
\biggl[\frac{\frac{1}{2}-\frac{2Sin^2\theta_{W}}{3}}{Cos\theta_W Sin\theta_W} -\frac{2Sin\theta_W}{3Cos\theta_W} \biggr]; \,\,\,\,
K_{13} = \frac{16e^3}{9}; \nn \\
K_{14} = \frac{4e^3}{3}\cdot \frac{Cos\theta_W}{Sin\theta_W} \cdot
\biggl[\frac{\frac{1}{2}-\frac{2Sin^2\theta_{W}}{3}}{Cos\theta_W Sin\theta_W} -\frac{2Sin\theta_W}{3Cos\theta_W} \biggr]; \,\,\,\,
K_{15} = \frac{8e}{3} M^2_W \frac{G_F}{\sqrt {2}} |V_{ud}|^2.
\label{A4}
\ea
where the quantities $M_W$, $e = \sqrt {4\pi\alpha}$, $G_F$, $V_{ud}$, and $\theta_W$ are the W-boson mass, the elementary electric charge ($\alpha \sim 1/137$),
the Fermi coupling constant, one of the elements of the Cabibbo - Kobayashi - Maskawa matrix, and the weak mixing angle, respectively. \\
Full square matrix element can be written as follows
\ba
\overline{|{\mathcal M}|^2} = \sum |M_1+ M_2+ M_3+ M_4+M_5+ M_6+M_7+ M_8+M_9+  \nn \\
+M_{10}+M_{11}+M_{12}+M_{13}+M_{14}+M_{15}|^2.
\label{AM}
\ea
\subsection{The kinematics \label{kin}}
For the subprocesses (2) Mandelstam invariants we can be written in the following form:
\ba
s = (p_1 + p_2)^2 = (k_1 + k_2 + k_3)^2; \,\,\,t = (p_1 - k_3)^2 = (k_1 + k_2 - p_2)^2; \nn \\
u = (p_2 - k_3)^2 = (k_1 + k_2 - p_1)^2; \,\,\,q_1 = (p_1 - k_1)^2 = (k_2 + k_3 - p_2)^2; \nn \\
q_2 = (p_2 - k_2)^2 = (k_1 + k_3 - p_1)^2.
\label{A5}
\ea
We can define five additional invariant parameters, which will be expressed through the main invariant variables (2):
\ba
s_1 = (k_1 + k_2)^2 = (p_1 + p_2 - k_3)^2;\,\,\,\,t_1 = (p_1 - k_2)^2 = (k_1 + k_3 - p_2)^2;  \nn \\
u_1 = (p_2 - k_1)^2 = (k_2 + k_3 - p_1)^2; \,\,\,\,q_3 = (k_3 + k_1)^2 = (p_1 + p_2 - k_2)^2;  \nn \\
q_4 = (k_3 + k_2)^2 = (p_1 + p_2 - k_1)^2.
\label{A6}
\ea
Using \eqref{A5} and \eqref{A6}, the scalar products of 4-momenta in the reaction we can be expressed in terms of these invariants
\ba
2(p_1p_2) = s;\,\,\,2(p_1k_3) = -t;\,\,\,2(p_2k_3) = -u;\,\,\,2(p_1k_1) = M^2_W - q_1;\,\,\,
2(p_2k_2) = M^2_W - q_2;  \nn \\
2(p_1k_2) = s + t + q_1 - M^2_W;\,\,\,2(p_2k_1) = s + u + q_1 - M^2_W;\,\,\,
2(k_1k_3) = q_2 - q_1 - t; \nn \\
2(k_2k_3) = q_1 - q_2 - u;\,\,\,2(k_1k_2) = s + t + u - 2 M^2_W;
\label{A7}
\ea
The particles are on their mass shell, and by considering the fact that the real photon does not have mass, and, neglecting quark masses, we then obtain:
\ba
p_1^2 = p_2^2 =0, \,\,\,\,k_3^2 = 0, \,\,\,\,k_1^2 = k_2^2 = M^2_W.
\label{A8}
\ea
Additional invariants parameters \eqref{A6} can be expressed through invariant variables \eqref{A5}
\ba
s_1 = s + t + u;\,\,\,\,t_1 = 2M^2_W - s - t - q_1;\,\,\,\, u_1 = 2M^2_W - s - u - q_2; \nn \\
q_3 = M^2_W - q_1 + q_2 - t;\,\,\,\,q_4 = M^2_W + q_1 - q_2 - u.
\label{A9}
\ea
To simplify the calculations, we omit the quark mass $m_q$ due to the kinematic region under consideration:
\ba
s + s_1 + q_1 + q_2 + t_1 + u_1 = 4M^2_W;\,\,\,\,s + q_2 + t_1 = M^2_W + q_3;\,\,\,\,s_1 + q_2 + u_1 = 2M^2_W + t; \nn \\
q_2 + t - q_1 - q_3 = 2t - M^2_W;\,\,\,\,q_2  - q_1 - q_3 = t - M^2_W;\,\,\,\,t + u + q_3 + q_4 = 2M^2_W; \nn \\
s_1 + q_2 + u_1 = 2M^2_W - u;\,\,\,\, q_4 + u + q_2 = q_1 + M^2_W;\,\,\,\,s + q_2 + t_1 = M^2_W - q_3; \nn \\
s + q_1 + q_2 + t_1 + u_1 + t +u = 4M^2_W - s;\,\,\,\,s_1 + q_3 + q_4 = s+ 2M^2_W;\,\,\,\,\nn \\
s + q_1 + u_1 = q_4;\,\,\,\,q_3 + q_4 = 2M^2_W - t - u;\,\,\,\,s + t + u + q_1 + q_2 = 2M^2_W.
\label{A10}
\ea
It would be convenient to introduce the following variables:
\ba
\rho = \frac{4M^2_W}{s},\,\,\,x = \frac{s_1}{s},\,\,\,y = cos(\widehat{\vec {p}_1\vec{k}_3}),\,\,\,
\beta_x = \sqrt{1 - \frac{4M^2_W}{x s}}.
\label{A11}
\ea
The kinematic limits for the above expression are
\ba
\rho \leq x \leq 1,\,\,\, -1 \leq y \leq 1,
\label{A12}
\ea
where $y$ is the cosine of the angle between $p_1$ (quark in initial state) and $k_3$ (photon in the final state) in the quark-anti-quark center-of-mass system.  \\
Using the above expression \eqref{A11} and \eqref{A12} for the variables $t$ and $u$, we obtain the following form:
\ba
t = -\frac{s}{2}(1-x)(1-\cos\theta_k), \nn \\
u = -\frac{s}{2}(1-x)(1+\cos\theta_k), \nn \\
\cos\theta_k = y.
\label{A13}
\ea
The completeness relation of the polarization vectors of photon can be written as
\ba
\sum_{\lambda}\varepsilon_{\mu}(k_3,\lambda)\varepsilon_{\nu}^{\ast}(k_3,\lambda) = -g_{\mu\nu}.
\label{A14}
\ea
The polarization vectors for massive spin-1 particles  such as the $Z$ and $W^{\pm}$ bosons, satisfy the completeness relation
\ba
\sum_{\lambda}\varepsilon_{\lambda}^{\mu}(k)\varepsilon_{\lambda}^{\nu\ast}(k) = -g_{\mu\nu} + \frac{k^{\mu}k^{\nu}}{M^2_W}.
\label{A15}
\ea
In the center-of-mass frame of the $W^+W^-$ system, we can parameterize the 4-vectors as $a = (a_0, a_z, a_x, a_y)$:
\ba
p_1 = E_1 (1, 0, 0, 1); \nn \\
p_2 = E_2 (1, 0, \sin\psi, \cos\psi); \nn \\
k_1 = E_{W^-}(1, \beta_x\sin\theta_1 \sin\theta_2, \beta_x\sin\theta_1 \cos\theta_2, \beta_x\cos\theta_1); \nn \\
k_2 = E_{W^+}(1, -\beta_x\sin\theta_1 \sin\theta_2, -\beta_x\sin\theta_1 \cos\theta_2, -\beta_x\cos\theta_1); \nn \\
k_3 = E_3 (1, 0, \sin\psi', \cos\psi').
\label{A16}
\ea
Using the conservation law of 4-momentum, we can determine the expression for $E_1, E_2, E_3$:
\ba
p_1 + p_2 - p_W - k_3 =0;  \nn \\
p_1 - p_W =  k_3 - p_2;
\label{A17}
\ea
here $p_W = \sqrt {s_1} = k_1 + k_2$ is the invariant mass of $W^+W^-$ pairs. If we take the square, then we get
\ba
-2E_1 \sqrt {s_1} + s_1 = -2p_2 k_3 = u;
\label{A18}
\ea
using the expression for $s_1$ from \eqref{A9},  we obtain the following expression for $E_1$:
\ba
E_1 = \frac{s+t}{2\sqrt {s_1}}.
\label{A19}
\ea
In an analogous method we will receive the same expression for $E_2$ and $E_3$:
\ba
E_2 = \frac{s+u}{2\sqrt {s_1}},\,\,\,\,\,\,E_3 = -\frac{t+u}{2\sqrt {s_1}}.
\label{A20}
\ea
Now we can determine $\cos\psi,\,\,\,\sin\psi,\,\,\,\cos\psi',\,\,\,\sin\psi'$. To do this, we need to use the following expression:
\ba
2p_1 p_2 = 2E_1 E_2 - 2|\vec {p}_1| |\vec {p}_2| \cos\psi.
\label{A21}
\ea
We get
\ba
\cos\psi = 1 - \frac{s}{2E_1 E_2}, \nn \\
\sin\psi = \sqrt {1-\cos^2\psi}.
\label{A22}
\ea
We must parametrize two more invariant variables $q_1$ and $q_2$:
\ba
q_1 = (p_1 - k_1)^2 = M^2_W - 2p_1k_1;  \nn \\
2p_1k_1 = 2E_1 E_{W^-} - 2|\vec {p}_1||\vec {k}_1|\cos\theta_1 = \frac{s+t}{2} -\frac{(s+t)}{2}\sqrt {1-\frac{4M^2_{W}}{xs}}\cos\theta_1 = \nn \\
= \frac{1}{2}(s+t)(1-\beta_x \cos\theta_1).
\label{A23}
\ea
We get
\ba
q_1 = M^2_W - \frac{1}{2}(s+t)(1-\beta_x \cos\theta_1),
\label{A24}
\ea
here $\beta_x = \sqrt {1-\frac{4M^2_{W}}{xs}}$. \\
Respectively we can parametrize $q_2$:
\ba
q_2 = (p_2 - k_2)^2 = M^2_W - 2p_2k_2;
\label{A25}
\ea
\ba
2p_2k_2 = s +u - 2p_2 k_1; \nn \\
2p_2k_1 = 2E_2 E_{W^-} - 2|\vec {p}_2||\vec {k}_1|\cos\theta_A = \frac{1}{2}(s+u)(1-\beta_x \cos\theta_A),
\label{A26}
\ea
Then for $2p_2k_2$ we obtain
\ba
2p_2k_2 = \frac{1}{2}(s+u)(1+\beta_x \cos\theta_A),
\label{A27}
\ea
where
\ba
\cos\theta_A = \sin\theta_1 \cos\theta_2 \sin\psi + \cos\theta_1 \cos\psi.
\label{A28}
\ea
Taking into account \eqref{A26}, \eqref{A27} and \eqref{A28} in \eqref{A25}, then for $q_2$ we obtain the following expression
\ba
q_2 = M^2_W - \frac{1}{2}(s+u)(1+\beta_x \cos\theta_A).
\label{A29}
\ea
Using formulas \eqref{A13} in \eqref{A19} and \eqref{A20},  for $E_1, E_2$ and $E_3$ we obtain a simple expression
\ba
E_1 = \frac{1}{4}\sqrt {s}\biggl(\frac{1}{\sqrt {x}}(1+y) + \sqrt {x} (1-y)\biggr); \nn \\
E_2 = \frac{1}{4}\sqrt {s}\biggl(\frac{1}{\sqrt {x}}(1-y) + \sqrt {x} (1+y)\biggr); \nn \\
E_3 = \frac{s}{\sqrt {s_1}}y(1-x).
\label{A30}
\ea
\subsection{The phase volume calculation for the  $q(p_1) + \bar {q}(p_2) \rightarrow W^+(k_1) + W^-(k_2) + \gamma (k_3)$ process
\label{phase}}

The phase volume of this process has the standard form:
\ba
d\Gamma_3(p_1, p_2, k_1, k_2, k_3) = (2\pi)^4 \delta^4(p_1 + p_2 - k_1 - k_2 - k_3)
\frac{d^3\vec{k}_1}{(2\pi)^3 2E_{W^-}} \frac{d^3\vec{k}_2}{(2\pi)^3 2E_{W^+}} \frac{d^3\vec{k}_3}{(2\pi)^3 2E_3},\,\,\,
\label{A31}
\ea
A derivation begins with introducing a unity factor:
\ba
1 = \int ds_1 \cdot \int \frac{d^3\vec{p}_W}{2E_W} \delta^4(p_W - k_1 - k_2), \nn \\
E^2_W = \vec{p}^2_W + s_1; \,\,\,\,\,\,s_1 = p^2_W = (k_1 + k_2)^2,
\label{A32}
\ea
and after resorting it is:
\ba
d\Gamma_3(p_1, p_2, k_1, k_2, k_3) = \int ds_1 \cdot \int
\frac{d^3\vec{k}_1}{(2\pi)^3 2E_{W^-}} \frac{d^3\vec{k}_2}{(2\pi)^3 2E_{W^+}} \frac{d^3\vec{k}_3}{(2\pi)^3 2E_3} \cdot \int \frac{d^3\vec{p}_W}{2E_W} \cdot \nn \\
\cdot (2\pi)^4 \delta^4(p_1 + p_2 - p_W - k_3)\cdot \delta^4(p_W - k_1 - k_2) \cdot
\label{A33}
\ea
Integrating over photon momenta removes the $\delta$-function, and using the properties of the $\delta$-function, we can rewrite:
\ba
\frac{d^3\vec{k}_3}{2E_3} \delta^4(p_1 + p_2 - p_W - k_3) = \delta (s + M^2_W -2\sqrt {s} E_W),
\label{A34}
\ea
If we conduct similar integration for momenta $k_2$, we get
\ba
\frac{d^3\vec{k}_2}{2E_{W^+}} \delta^4(p_W - k_1 - k_2) = \delta (M^2_{W^+} - M^2_{W^-} - p^2_W + 2\sqrt {s_1} E_{W^-});
\label{A35}
\ea
From this $\delta$-function one can obtain an expression for the energy of $W^-$boson
\ba
E_{W^-} = \frac{s_1 - M^2_{W^+} + M^2_{W^-}}{2\sqrt {s_1}} = \frac{1}{2}\sqrt {s_1}, \,\,\,\,\, p^2_W = s_1.
\label{A36}
\ea
In an analogous form it is possible to derive an expression for the energy of $W^+$ boson:
\ba
E_{W^+} = \frac{s_1 + M^2_{W^+} - M^2_{W^-}}{2\sqrt {s_1}} = \frac{1}{2}\sqrt {s_1}.
\label{A37}
\ea
After integrating over the momenta $k_3$ and $k_2$, we obtain an expression for $d\Gamma_3$:
\ba
d\Gamma_3 \sim \frac{d^3\vec{k}_1}{2E_{W^-}} \cdot \frac{d^3\vec{p}_W}{2E_W} \delta (s + M^2_W -2\sqrt {s}E_W)\cdot \delta (M^2_{W^+} - M^2_{W^-} - s_1 + 2\sqrt {s_1} E_{W^-}).
\label{A38}
\ea
In a analogous method, we obtain the following expression after integration over $k_1$:
\ba
\frac{d^3\vec{k}_1}{2E_{W^-}} \cdot \delta (M^2_{W^+} - M^2_{W^-} - s_1 + 2\sqrt {s_1} E_{W^-}) = \frac{k_1^{W^-}}{4 \sqrt {s_1}} d\Omega_{W^-}.
\label{A39}
\ea
As a result, we get
\ba
\frac{d^3\vec{p}_W}{2E_W} \cdot \frac{k_1^{W^-}}{4 \sqrt {s_1}} d\Omega_{W^-}\cdot \delta (s + M^2_W -2\sqrt {s} E_W) =
\frac{p_W k_1^{W^-}}{16 \sqrt {s} \sqrt {s_1}}d\Omega_{W} d\Omega_{W^-},
\label{A40}
\ea
here $k_1^{W^-} = \frac{1}{2}\sqrt {x s} \beta_x.$  \\
If this expression is put in \eqref{A32}, then we get
\ba
d\Gamma_3(p_1, p_2, k_1, k_2, k_3) = (2\pi)^{-5} \int ds_1 \cdot \frac{p_W k_1^{W^-}}{16 \sqrt {s} \sqrt {s_1}}d\Omega_{W} d\Omega_{W^-}.
\label{A41}
\ea
Finally, by integrating over the angle and after simplifying, we have
\ba
d\Gamma_3 = \frac{s \beta_x}{512 \pi^4} (1-x) dx dy d(\cos\theta_1) d\theta_2.
\label{A42}
\ea
Here, the angles $\theta_1$ and $\theta_2$ range from 0 to $\pi$.
\subsection{The transverse momentum and rapidity distributions}
\ba
\hat {s} = x_1 x_2 S; \nn \\
q_1 = (p_1 - k_1)^2 = M^2_W - \frac{S}{2}x_1 x_{TW^-}e^{-y_1} =
M^2_W - x_1 \sqrt {s} \sqrt {p^2_{tW^-} + M^2_W}e^{-y_1}, \nn \\
q_2 = (p_2 - k_2)^2 = M^2_W - \frac{S}{2}x_2 x_{TW^+}e^{-y_2} =
M^2_W - x_2 \sqrt {s} \sqrt {p^2_{tW^+} + M^2_W}e^{-y_2}.
\label{A43}
\ea
In the above expression, $S$ is the square of the hadronic center of mass energy, $y_1$ and $y_2$ are rapidities of $W^-$ and $W^+$ bosons, respectively,
where $p_{tW^-}$ and $p_{tW^+}$ are transverse momenta of $W^-$ and $W^+$ bosons, respectively. \\
The kinematic limits for the above expression are
\ba
\frac{\sqrt {p^2_{tW^+} + M^2_W}}{\sqrt {s} - \sqrt {p^2_{tW^-} + M^2_W}e^{y_1}} \leq y_2 \leq \frac{\sqrt {s}-\sqrt {p^2_{tW^-} + M^2_W}e^{y_1}}{\sqrt {p^2_{tW^+} + M^2_W}}.
\label{A44}
\ea
\ba
y_{1max} = -y_{1min} = min\left\{Y_1;\,\, \cosh^{-1}\left(\frac{\sqrt {s}}{2\sqrt {p^2_{tW^-} + M^2_W}}\right); \,\,\ln\left(\frac{\sqrt {s}-\sqrt {p^2_{tW^+} + M^2_W}e^{-Y_2}}{\sqrt {p^2_{tW^-} + M^2_W}}\right)\right\}, \,\,\,\,\,
\label{A45}
\ea
in the numerical applications we take $Y_2$ =2.5.  \\
Using the lowest order kinematics described previously, another variable that is often used in studies
of pair (or jet) production is the pair (dijet) invariant mass $M^2_{W^+W^-}$. This is to be given by
\ba
M^2_{W^+W^-} = s_1 = 2M^2_W + 2 \sqrt {p^2_{T_{W^-}} + M^2_W} \sqrt {p^2_{T_{W^+}} + M^2_W} [1+ cosh (y_1 - y_2)].
\label{A46}
\ea

The total cross section for the $pp \rightarrow W^+W^-\gamma$ process at the hadronic level obtained by convoluting $\hat {\sigma}_{ab}$
with parton distribution functions (PDFs) of the colliding protons and four-momenta $P_1$ and $P_2$ can be written in the following form:
\ba
\sigma (P_1, P_2) &=& \sum_{a,b} \frac{1}{1+\delta_{ab}} \int\limits_{x_{1,min}}^1 dx_1 \int\limits_{x_{2,min}}^1 dx_2 \cdot f_{a/A} (x_1, \mu^2_f) f_{b/B}(x_2,\mu^2_f) \hat {\sigma}_{ab}(x_1, x_2, p_1, p_2, M^2_W, \mu^2_f) + \nn \\
&&+(A \leftrightarrow B),
\label{A47}
\ea
where the sum is over all partons that contribute to the process (a runs over the partons of proton A, and b
runs over the partons of proton B);
$x_1$ and $x_2$ are the fractional momenta of the partons participating in the fundamental subprocesses,
and $m$ is the quark mass, $M_W$ is the mass of $W$ boson, the variable $\mu$ is the factorization scale,
and the indices $a$ and $b$ specify the types of the incoming partons;
$f_{a/A}$ and $f_{b/B}$ are parton distribution functions (PDFs) for the partons
$a$, $b$, in the proton $A$, $B$;
$\hat {\sigma}_{ab}$ is the partonic cross section for interactions of two partons.
The Kronecker delta factor necessary for identical initial-state partons. \\
For the parton distribution functions (PDFs) we use the sets of MSTW2008 parametrization \cite{MRSW}.

From kinematics, we can determine $x_1$ and $x_2$ in the following form
\ba
x_1 =
\biggl(x_2 \sqrt {s} \sqrt {p^2_{tW^-} + M^2_W}e^{y_1} + x_2 \sqrt {s} \sqrt {p^2_{tW^+} + M^2_W}e^{-y_2} - \nn \\
-2\sqrt {s} \sqrt {p^2_{tW^-} + M^2_W} \sqrt {p^2_{tW^+} + M^2_W} (1+ e^{y_1 - y_2})\biggr) /\biggl(x_2 s -\sqrt {s} \sqrt {p^2_{tW^-} + M^2_W}e^{y_1} - \nn \\
-\sqrt {s} \sqrt {p^2_{tW^+} + M^2_W}e^{-y_2}\biggr),
\label{A48}
\ea
$x_1$ gets a minimum at $x_2=1$, that is
\ba
x_{1, min} =
\biggl(\sqrt {s} \sqrt {p^2_{tW^-} + M^2_W}e^{y_1} + \sqrt {s} \sqrt {p^2_{tW^+} + M^2_W}e^{-y_2} - \nn \\
-2\sqrt {s} \sqrt {p^2_{tW^-} + M^2_W} \sqrt {p^2_{tW^+} + M^2_W} (1+ e^{y_1 - y_2})\biggr) /\biggl(s -\sqrt {s} \sqrt {p^2_{tW^-} + M^2_W}e^{y_1} - \nn \\
-\sqrt {s} \sqrt {p^2_{tW^+} + M^2_W}e^{-y_2}\biggr),
\label{A49}
\ea
and
\ba
x_{2, min} =
\biggl(x_1 \sqrt {s} \sqrt {p^2_{tW^-} + M^2_W}e^{y_1} + x_1 \sqrt {s} \sqrt {p^2_{tW^+} + M^2_W}e^{-y_2} - \nn \\
-2\sqrt {s} \sqrt {p^2_{tW^-} + M^2_W} \sqrt {p^2_{tW^+} + M^2_W} (1+ e^{y_1 - y_2})\biggr) /\biggl(x_1 s -\sqrt {s} \sqrt {p^2_{tW^-} + M^2_W}e^{y_1} - \nn \\
-\sqrt {s} \sqrt {p^2_{tW^+} + M^2_W}e^{-y_2}\biggr).
\label{A50}
\ea
The total cross sections for the partonic process $q + \bar {q} \to W^+ + W^- + \gamma$ have the form
\ba
\hat {\sigma}_{ab} = \frac{1}{2\hat {s}}\frac{1}{4N^2_C}\int d\Gamma_3 \sum_{\substack {spins, \\ color}} \overline{|\mathcal M_{ab}|^2},
\label{A51}    
\ea
where $d\Gamma_3$ is the three-body phase space element, the summation is taken over the spins and colors of the initial and final states,
$\hat {s}$ is the centre-of-mass energy-squared of the parton-parton collision, $N_C = 3$ is the number of colors in QCD.  \\
Now we want to derive an expression for the differential cross section,
which will depend on the transverse momenta and the rapidity of the $W^+W^-$ bosons in the final state.  \\
To this end, we use \eqref{A43} in the expression for the square of the amplitude \eqref{AM}, and using \eqref{A47}
we can theoretical by calcuclate the cross sections according to
\ba
\frac{d\sigma}{dp_{T,W^+} dp_{T,W^-}} &=& \int dy_1 \int dy_2 \cdot \tau \frac{d\sigma}{dp_{T,W^+} dp_{T,W^-} dy_1 dy_2},
\label{A54}
\ea
where $\tau = \frac{\hat s}{S} = x_1 x_2$.  \\
The cross section distribution of the rapidities, we obtain the following formulas
\ba
\frac{d\sigma}{dy_1 dy_2} &=& \int dp_{T,W^+} \int dp_{T,W^-} \cdot \frac{16}{S} p_{T,W^+}p_{T,W^-}\cdot \frac{d\sigma}{dp_{T,W^+} dp_{T,W^-}dy_1 dy_2},
\label{A55}
\ea
the $y_1$ and $y_2$ integrations limits are given by \eqref{A44} and \eqref{A45}.
Integration limits for the $p_{T, W}$, we obtain the following expression:
\ba
p_{T, W min}^{exp} = p_{T, min}^{exp} >0, \nn \\
p_{T, W max} = \sqrt {\frac{S}{Exp [2y_1] + 2Exp [y_1-y_2]+ Exp [-2y_2]}-M^2},
\ea
for the case $y_1 = y_2 = 0$, for $p_{T, W max} $ we obtain
\ba
p_{T, W max} = \frac{1}{2}\sqrt{S-4M_W^2}.
\ea
\section{Numerical results}
\label{NR}
\begin{figure}[!htb]
       \centering
       \includegraphics[width=0.72\linewidth]{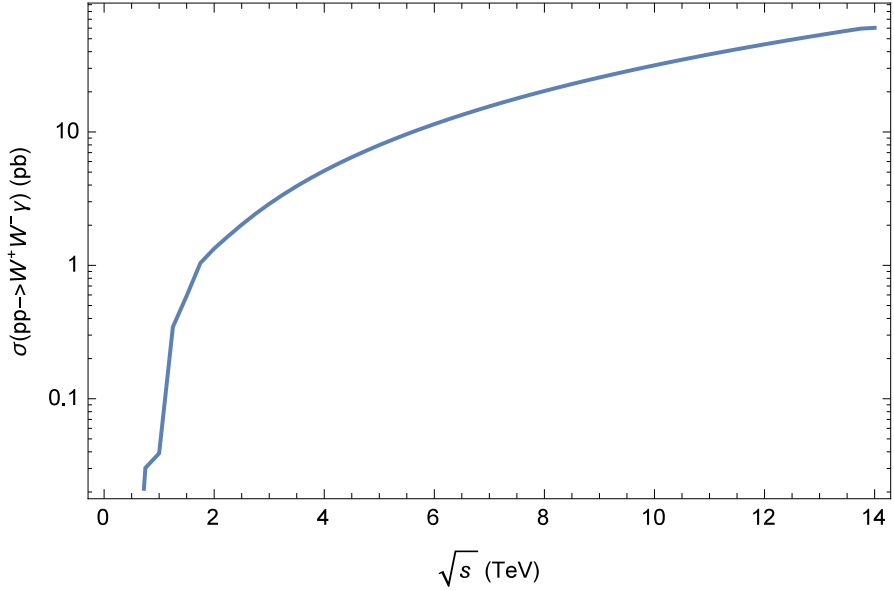}
       \caption{The total cross-section for the $W^+W^-\gamma$ production process as a function of the center-of-mass energy $\sqrt s$.}
       \label{CS}
\end{figure}

In this Section, we present numerical results by explicitly considering the distribution of the total cross-section,
the transverse-momentum distribution, and the rapidity distribution of $W^+W^-$ bosons in proton-proton collisions at LHC energy. \\
We calculate the total cross-section by formulas \eqref{A47} for $WW\gamma$ production processes as a function of collider centre-of-mass (CM) energy $\sqrt s$
ranging from 0.5 to 14 TeV.
The obtained result on total cross-cestion is shown in Fig.~\ref{CS}.
The central values of renormalization ($\mu_R$), and factorization ($\mu_F$) scales are set to $\mu_R = \mu_F = M_W$.

\begin{figure}[!htb]
       \includegraphics[width=0.48\linewidth]{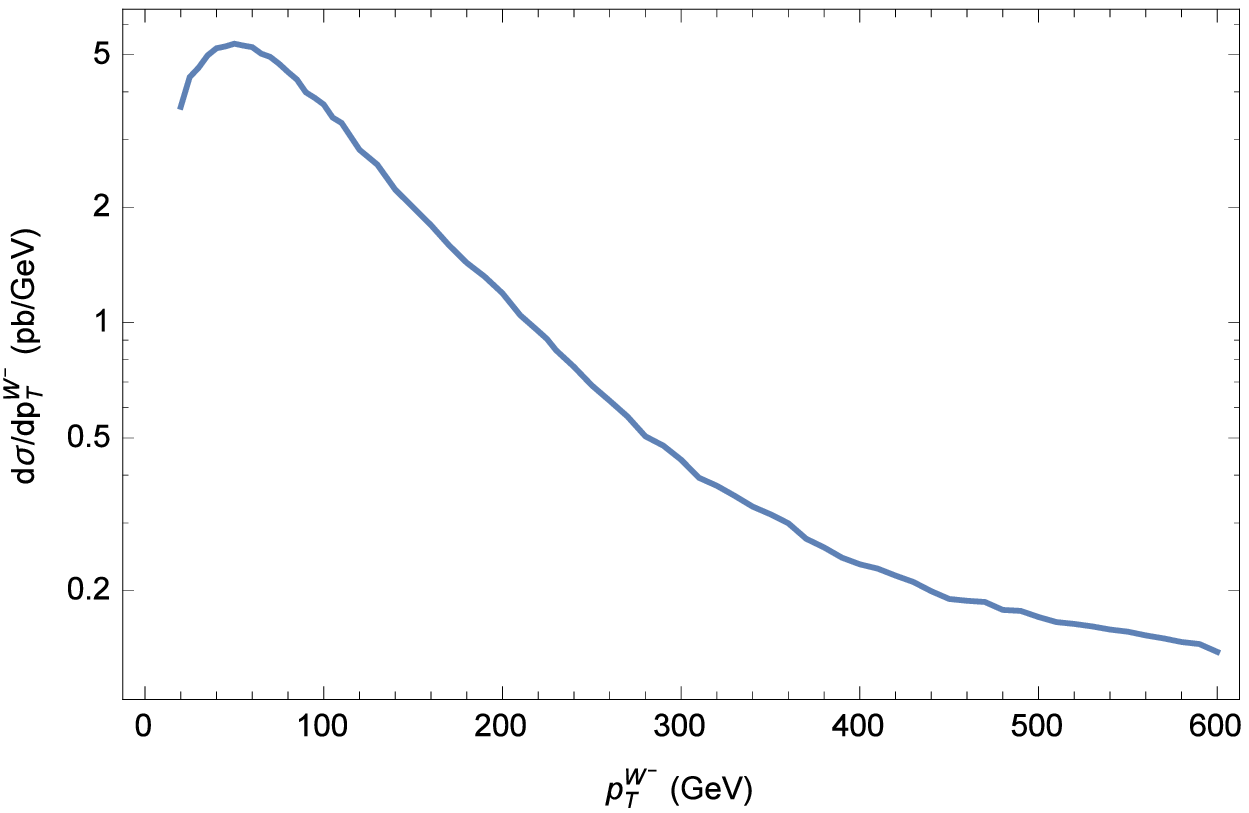}
       \includegraphics[width=0.48\linewidth]{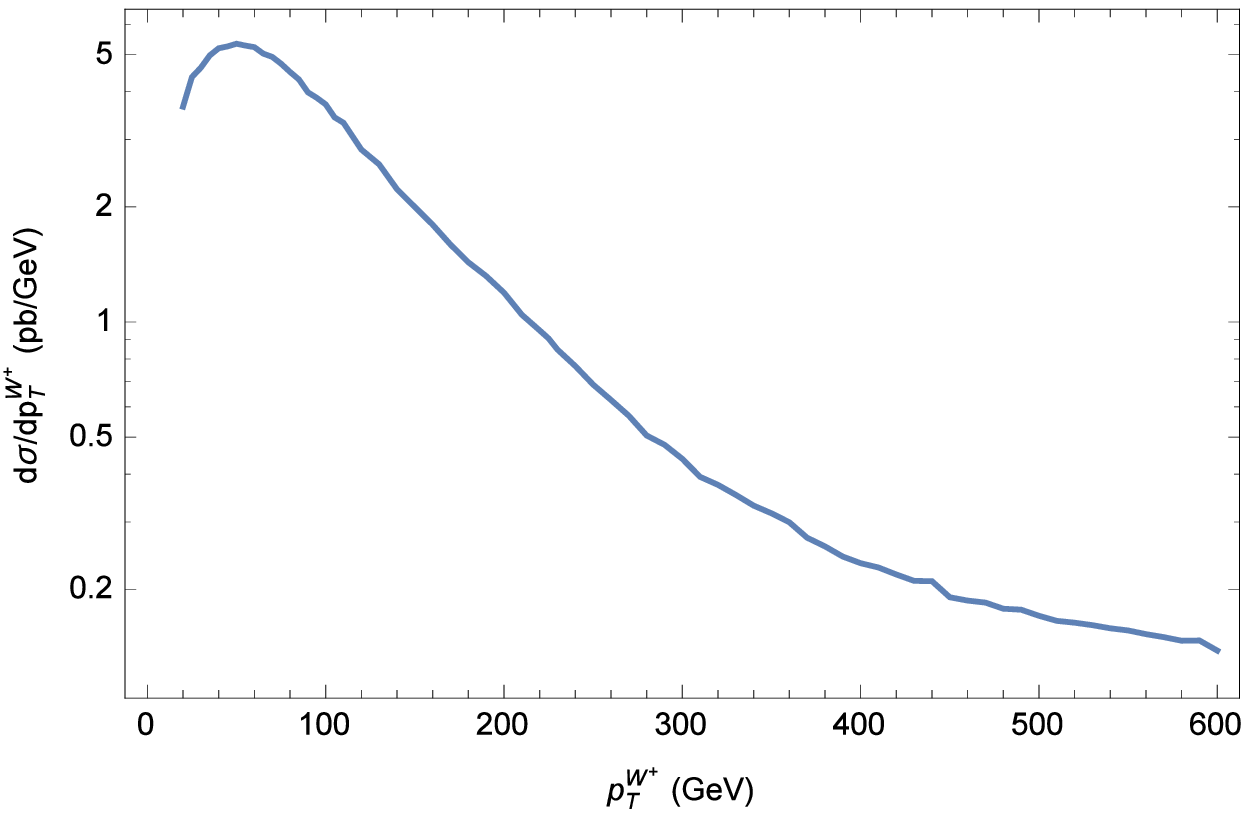}
       \caption{$W^+ \,\,{\textrm and}\,\, W^-$ transverse momentum distributions at the center-of-mass energy $\sqrt {s} = 14 \,\,TeV$ in $pp$ interactions.}
       \label{fig5}
\end{figure}

\begin{figure}[!htb]
       \centering
       \includegraphics[width=0.72\linewidth]{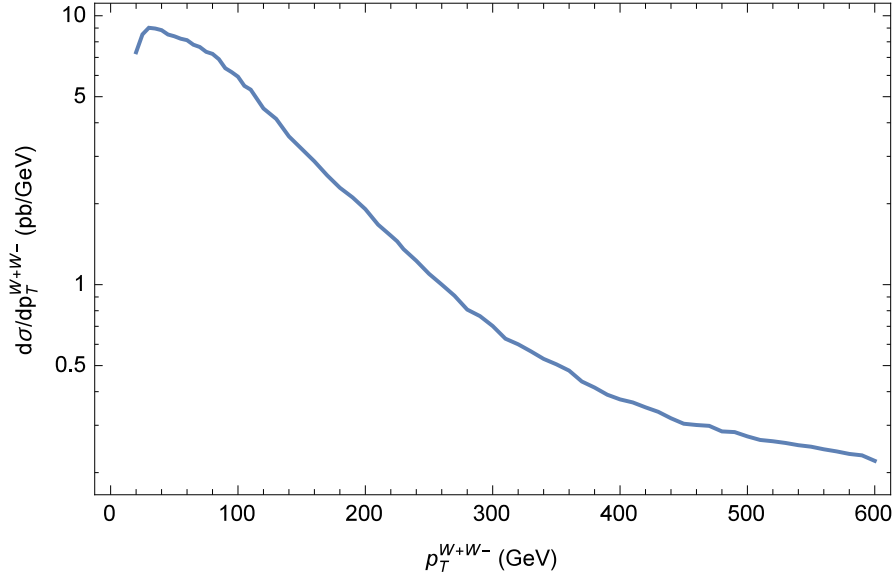}
       \caption{Transverse momentum distribution of the $W^+W^-$ pair at the center-of-mass energy $\sqrt {s} = 14 \,\,TeV$ in $pp$ interactions.}
       \label{fig6}
\end{figure}

\begin{figure}[!htb]
       \centering
       \includegraphics[width=0.72\linewidth]{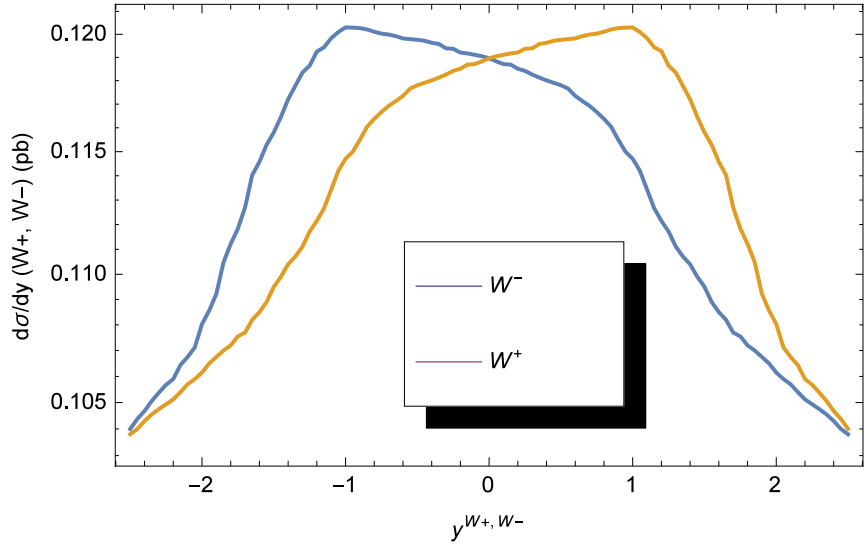}
       \caption{The rapidity distributions of the $W^+$ and $W^-$ bosons at the center-of-mass energy $\sqrt {s} = 14 \,\,TeV$ in $pp$ interactions:
       in the left peak ($W^-$), and  in the right peak ($W^+$), respectively.}
       \label{fig7}
\end{figure}

\begin{figure}[!htb]
       \centering
       \includegraphics[width=0.72\linewidth]{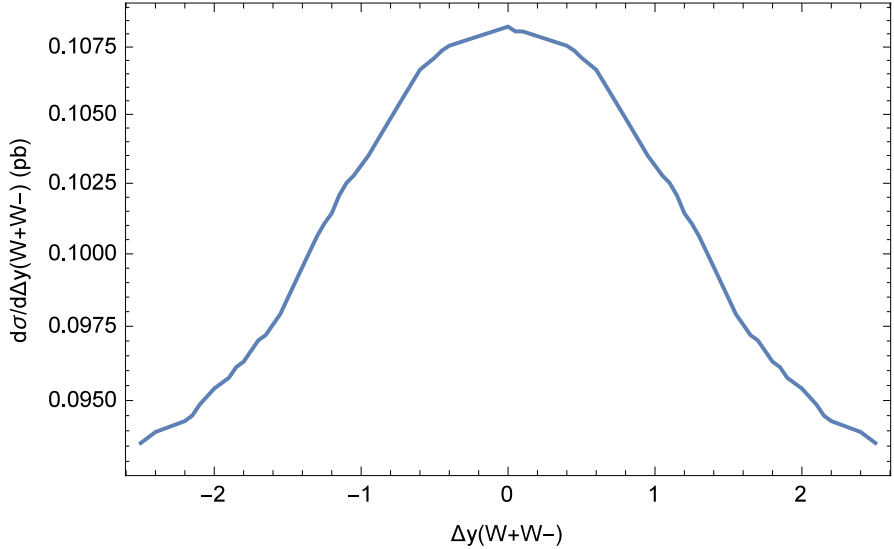}
       \caption{Distribution of the correlation of rapidity between $W^+$ and $W^-$ bosons: $\Delta y(W+W-) = y^{W+} - y^{W-}$, at the center-of-mass energy $\sqrt {s} = 14 \,\,TeV$.}
       \label{fig8}
\end{figure}

We calculated the transverse momentum ($p_T$) distributions in the region from 0.5 to 14 TeV of $W^+$ boson (for the $W^-$ boson transverse momentum distribution is analogous)
in the final state for $pp \to W^+W^-\gamma$ process of the $\sqrt {s} = 14\,\,TeV$.
The results of our numerical calculations have been represented in Fig.~\ref{fig5}.  \\
We now present the inclusive transverse-momentum spectrum of the $W$ vector-boson pair.
We calculated the differential cross section for the transverse momentum ($p_T (WW)$) distribution of the  $W$ boson pair in the regions
$p_T = 20 \div 600 \,\,GeV$ at the $\sqrt {s} = 14\,\, TeV.$  We show our results of the transverse momentum distribution of the $W$- boson pair in Fig.~\ref{fig6}.\\
It can be predicted that the feature of the $p_T$ distributions at the LHC characterize as searches in the triple gauge boson measurements.  \\
We will now study the distribution of the differential cross section on the rapidities ($y$) of $W^+$ and $W^-$ bosons in the final state.
In Fig.~\ref{fig7}, we present the rapidity ($y$) distributions, in the reqions $-2.5 \leq y \leq 2.5$ of the final $W^+$ and $W^-$ bosons
for the $pp \to W^+W^-\gamma$ process at $\sqrt {s} = 14\,\,TeV$.
Also, we have calculated the distribution of the differential cross section for correlation of the rapidities between $W^+$ and $W^-$ bosons.
The distribution of the correlation of rapidities is depicted in Fig.~\ref{fig8}
\section{Conclusion}
\label{Conclusion}
In the present paper, we have studied the transverse momentum and rapidity distributions of vector bosons in proton - proton collisions. \\
We studied the $pp \rightarrow W^+W^-\gamma$ process using the quark (and anti-quark) distribution functions in the initial
state set of the MSTW2008 \cite{MRSW} parametrization in the framework of the Standard Model,
which are intensively investigated in ATLAS and CMS at the Large Hadron Collider. \\
We have presented the numerical results for $W^+W^-\gamma$ production at LHC energy.  \\
Having all the tools to calculate and using equation \eqref{A47} together with the hard-scattering amplitude in equation \eqref{AM}
we performed the parton level calculations.

We estimated the cross section of the $W$ bosons production as a function of transverse momentum $p_T$ and rapidity $y$ of the $W$
bosons. The obtained results using the set of uPDFs \cite{MRSW} can be used for such phenomenological studies.
We used the experimental cuts of the transverse momentum and rapidity of the $W$ bosons used by the ATLAS and CMS
experiments in their measurements, which are available to leading transverse momentum of lepton $p_T > 20 \,\,\,GeV$ and
absolute value of rapidity $|\eta| < 2.5$. \\
In this present paper, we investigated the distribution of the transverse momentum ($p_T$) and the rapidity ($y$) of $W^+$ and $W^-$ bosons separately,
and also the distribution of the total cross section from the centre-of-mass energy $\sqrt {s}$ at LHC energy. \\
In this paper, we studied several cases of phenomenological interest, which demonstrate the effect of NLO corrections.  \\
It can be predicted that the production of WW boson pair and $WW\gamma$ production is an important source of both the background for
Higgs and the search for new physics at LHC.


\end{document}